\documentclass[aps,prd,onecolumn,groupedaddress,showpacs,nofootinbib,amssymb]{revtex4-2}
\usepackage[dvips]{graphicx}
\usepackage{amssymb}
\usepackage{amsmath}
\usepackage{graphicx,,color}
\usepackage{amsfonts}
\usepackage{bm}
\usepackage{cancel}
\usepackage{comment}
\usepackage{floatflt}
\usepackage{slashed}

\newcommand\be{\begin{equation}}
\newcommand\ee{\end{equation}}

\allowdisplaybreaks[4]

\begin{document}

\title{A Stiff Pre-CMB Era with a Mildly Blue-tilted Tensor Inflationary Era can Explain the 2023 NANOGrav Signal}
\author{V.K. Oikonomou,$^{1,2}$}\email{voikonomou@gapps.auth.gr;v.k.oikonomou1979@gmail.com}
\affiliation{$^{1)}$Department of Physics, Aristotle University of
Thessaloniki, Thessaloniki 54124, Greece \\ $^{2)}$L.N. Gumilyov
Eurasian National University - Astana, 010008, Kazakhstan}

 \tolerance=5000

\begin{abstract}
We examine the effects of a stiff pre-recombination era on the
present day's energy spectrum of the primordial gravitational
waves. If the background total equation of state parameter at the
pre-recombination era is described by a kination era one, this
directly affects the modes with characteristic wavenumbers which
reenter the Hubble horizon during this stiff era. The stiff era
causes a broken-power-law effect on the energy spectrum of the
gravitational waves. We use two approaches, one model agnostic and
a specific model that can realize this scenario. In all cases, the
inflationary era can be realized either by some theory leading to
a standard red-tilted tensor spectral index or by some theory
which has a mild tensor spectral index $n_{\mathcal{T}}=0.17-0.37$
like an Einstein-Gauss-Bonnet theory. For the model agnostic
scenario case, the NANOGrav signal can be explained by the stiff
pre-recombination era combined with an inflationary era with a
mild blue-tilted tensor spectral index $n_{\mathcal{T}}=0.37$ and
a low-reheating temperature $T_R\sim 0.1$GeV. In the same case,
the red-tilted inflationary theory signal can be detectable by the
future LISA, BBO and DECIGO experiments. The model dependent
approach is based on a Higgs-axion model which can yield multiple
deformations of the background total equation of state parameter,
causing multiple broken-power-law behaviors occurring in various
eras before and after the recombination era. In this case, the
NANOGrav signal is explained by this model in conjunction with  an
inflationary era with a really mild blue-tilted tensor spectral
index $n_{\mathcal{T}}=0.17$ and a low-reheating temperature
$T_R\sim 20\,$GeV. In this case, the signal can be detectable by
the future Litebird experiment, which is a very characteristic
pattern in the tail of the primordial gravitational wave energy
spectrum.
\end{abstract}


\maketitle

\section{Introduction}

Inflation \cite{inflation1,inflation2,inflation3,inflation4} is
now put to test from most current and forthcoming experiments. The
aim is to pinpoint the B-modes directly experimentally, and this
is the central focus of the stage 4 Cosmic Microwave Background
(CMB) experiments \cite{CMB-S4:2016ple,SimonsObservatory:2019qwx}.
Apart from the B-modes, there are indirect ways to detect an
inflationary era, for example the detection of a small (or
negligible) anisotropy stochastic gravitational wave background
\cite{Hild:2010id,Baker:2019nia,Smith:2019wny,Crowder:2005nr,Smith:2016jqs,Seto:2001qf,Kawamura:2020pcg,Bull:2018lat,LISACosmologyWorkingGroup:2022jok}.
These future gravitational wave experiments will directly probe
the inflationary tensor perturbations. Apart from the LIGO-Virgo
successes and exciting observations, recently the NANOGrav
collaboration confirmed the well anticipated stochastic
gravitational wave detection \cite{nanograv}, which was also
confirmed by other pulsar timing arrays (PTA) experiments
\cite{Antoniadis:2023ott,Reardon:2023gzh,Xu:2023wog}. The
scientific community was almost certain that the 2020 signal
detection was not due to the actual pulsar red noise but due to a
stochastic gravitational wave background, which was confirmed in
2023 due to the presence of Hellings-Downs correlations. After the
2023 NANOGrav announcement, a large stream of research articles
were produced, trying to explain the signal, for the cosmological
perspective see for example
\cite{sunnynew,Oikonomou:2023qfz,Cai:2023dls,Han:2023olf,Guo:2023hyp,Yang:2023aak,Addazi:2023jvg,Li:2023bxy,Niu:2023bsr,Yang:2023qlf,Datta:2023vbs,Du:2023qvj,Salvio:2023ynn,Yi:2023mbm,You:2023rmn,Wang:2023div,Figueroa:2023zhu,Choudhury:2023kam,HosseiniMansoori:2023mqh,Ge:2023rce,Bian:2023dnv,Kawasaki:2023rfx,Yi:2023tdk,An:2023jxf,Zhang:2023nrs,DiBari:2023upq,Jiang:2023qbm,Bhattacharya:2023ysp,Choudhury:2023hfm,Bringmann:2023opz,Choudhury:2023hvf,Choudhury:2023kdb,Huang:2023chx,Jiang:2023gfe,Zhu:2023lbf,Ben-Dayan:2023lwd,Franciolini:2023pbf,Ellis:2023oxs,Liu:2023ymk,Liu:2023pau,Madge:2023cak,Huang:2023zvs,Fu:2023aab,Maji:2023fhv,Gangopadhyay:2023qjr},
see also
\cite{Schwaller:2015tja,Ratzinger:2020koh,Ashoorioon:2022raz,Choudhury:2023vuj,Choudhury:2023jlt,Choudhury:2023rks,Bian:2022qbh}
for earlier works in this perspective, and also
\cite{Guo:2023hyp,Yang:2023aak,Machado:2018nqk} for the axion
description. Although it is not certain whether the 2023 NANOGrav
signal has a cosmological or astrophysical source, or even a
combination of the two, there are theoretical hints toward the
cosmological description. For the astrophysical description of the
signal, see the recent review \cite{Regimbau:2022mdu}. Currently
there are some issues that render the cosmological description
more plausible, although in principle the signal may be a hybrid
of astrophysical and cosmological origin. The obstacles for the
astrophysical description are firstly the lack of a concrete
solution for the last parsec problem \cite{Sampson:2015ada},
secondly, the absence of large anisotropies in the NANOGrav signal
\cite{NANOGrav:2023gor} and thirdly the complete absence of
isolated supermassive black hole merger events. Clearly, the
detection of large anisotropies in the future may point out that
the astrophysical description is plausible
\cite{Sato-Polito:2023spo,Ding:2023xeg}. However, some hint in
this direction should be present even in the 2023 data, so the
near future will shed light on these issues. On the other hand,
there are a lot of cosmological scenarios that may explain the
signal, like phase transitions, cosmic strings, primordial black
holes and so on. With regard to the inflationary perspective, it
seems that ordinary inflation cannot explain in any way the 2023
NANOGrav signal, unless a highly blue-tilted tensor spectral index
is produced and a significantly low-reheating temperature is
required
\cite{sunnynew,Oikonomou:2023qfz,Benetti:2021uea,Vagnozzi:2020gtf}.
In principle, a blue-tilted tensor spectral index can be generated
in the context of many theories, like string gas theories
\cite{Kamali:2020drm,Brandenberger:2015kga,Brandenberger:2006pr},
or some Loop Quantum Cosmology scenarios can also predict such a
tensor spectrum
\cite{Ashtekar:2011ni,Bojowald:2011iq,Mielczarek:2009vi,Bojowald:2008ik},
and moreover the non-local version of Starobinsky inflation
\cite{Calcagni:2020tvw,Koshelev:2020foq,Koshelev:2017tvv} which
also yield an acceptable amount of non-Gaussianities. Furthermore,
it is important to note that conformal field theory can yield a
blue-tilted tensor spectrum \cite{Baumgart:2021ptt}, see also
\cite{Cai:2020qpu} for a different perspective. From the
aforementioned examples, the only which yields a strongly
blue-tilted tensor spectral index is the non-local Starobinsky
model, however in such a case a low reheating temperature is also
needed. In the case that the reheating temperature is actually too
low, the electroweak phase transition is put into peril, since it
cannot be realized thermally, it is required that the reheating
temperature is at least 100$\,$GeV. We shall discuss this
important issue at a later section. Thus the situation is somewhat
perplexed since it is not certain what causes the 2023 NANOGrav
signal. Only the combination of data coming from all the future
experiments like LISA and the Einstein Telescope, including
NANOGrav, may point out toward to a specific model behind the
stochastic signal. This synergy between experiments may also help
determining the reheating temperature.

In view of the above line of research, in this work we shall
report an intriguing result in the context of inflationary
theories and their ability to explain the 2023 NANOGrav signal.
Our analysis requires a stiff, or nearly stiff pre-CMB era, which
can be caused by various mechanisms, see for example
\cite{Co:2021lkc,Gouttenoire:2021jhk,Giovannini:1998bp,Oikonomou:2023bah,Ford:1986sy,Kamionkowski:1990ni,Grin:2007yg,Visinelli:2009kt,Giovannini:1999qj,Giovannini:1999bh,Giovannini:1998bp}.
The approach of Ref. \cite{Co:2021lkc} remarkably fits the line of
research we shall adopt in this work, and the pre-CMB stiff or
kination era is caused by the axion. Actually the authors perform
a thorough and scientifically sound analysis to put constraints on
such a pre-CMB kination era caused by the axion. We shall not use
a specific model in our work, however the constraints of
\cite{Co:2021lkc} are taken into account in our analysis. In a
previous work we considered the effects of stiff eras caused by
axions during reheating \cite{Oikonomou:2023bah}, so this work is
basically in the same spirit as our previous one. Also it is
noticeable that recently another work which invokes a stiff era
and the possibility of explaining the NANOGrav 2023 signal
appeared in the literature \cite{Harigaya:2023pmw}, in a different
approach though, compared to the one we shall present in this
work. Coming back to our case, the stiff era is required to
correspond at a wavenumber range $k=0.4-0.9\,$Mpc$^{-1}$, so just
before the recombination era\footnote{The stiff era is assumed to
occur exactly before the CMB modes ($k\sim 0.002\,$Mpc$^{-1}$)
reentered the horizon so well after the matter-radiation equality.
}, and recall that the CMB pivot scale is at
$k=0.002$$\,$Mpc$^{-1}$. Note that the CMB modes with
$k=0.002$$\,$Mpc$^{-1}$ reentered the horizon at $z=1100$ so
during recombination, so the stiff era is assumed to have occurred
before the recombination so when $k=[0.4,0.9]\,$Mpc$^{-1}$ hence
for a redshift $2500>z>1100$ and and well below the
matter-radiation redshift. The inflationary era can be described
by a standard red-tilted or a blue-tilted theory. With regard to
the red-tilted inflationary era, it can be described by a standard
inflationary scenario or some modified gravity
\cite{reviews1,reviews2,reviews3,reviews4,reviews5,Sebastiani:2016ras,reviews6},
while the blue-tilted inflationary era can be described for
example by an Einstein-Gauss-Bonnet gravity. The striking new
result is that the tensor spectral index in the latter case is not
required to be strongly blue-tilted, but it can be of the order
$\sim \mathcal{O}(0.37)$, which can be easily generated by an
Einstein-Gauss-Bonnet theory. In the case of a blue-tilted
inflationary era, the compatibility with the NANOGrav result comes
easily when the reheating temperature is of the order $T\sim
\mathcal{O}(0.1)\,$GeV, while the red-tilted inflationary case
scenario is not detectable by the NANOGrav, however the prediction
is that can be either detected by LISA or by the Einstein
Telescope, not simultaneously though, depending on the reheating
temperature and the total equation of state (EoS) parameter
post-inflationary.

Regarding the stiff pre-CMB era, we shall assume an agnostic
approach without proposing a model for this, except for the last
section where we shall discuss a potentially interesting model.

\section{Stiff Pre-CMB era, Standard or Blue-tilted Inflation and Gravitational Waves}

Apparently, after the recent successes in gravitational wave
detections, the interest of theoretical physicists and
cosmologists is focused on gravitational waves and specifically
those having primordial origin. In the literature there exist
several works on primordial gravitational waves, for a mainstream
of research articles and reviews see
\cite{Kamionkowski:2015yta,Turner:1993vb,Boyle:2005se,Zhang:2005nw,Caprini:2018mtu,Clarke:2020bil,Smith:2005mm,Giovannini:2008tm,Liu:2015psa,Vagnozzi:2020gtf,Giovannini:2023itq,Giovannini:2022eue,Giovannini:2022vha,Giovannini:2020wrx,Giovannini:2019oii,Giovannini:2019ioo,Giovannini:2014vya,Giovannini:2009kg,Kamionkowski:1993fg,Giare:2020vss,Zhao:2006mm,Lasky:2015lej,
Cai:2021uup,Odintsov:2021kup,Lin:2021vwc,Zhang:2021vak,Visinelli:2017bny,Pritchard:2004qp,Khoze:2022nyt,Casalino:2018tcd,Oikonomou:2022xoq,Casalino:2018wnc,ElBourakadi:2022anr,Sturani:2021ucg,Vagnozzi:2022qmc,Arapoglu:2022vbf,Giare:2022wxq,Oikonomou:2021kql,Gerbino:2016sgw,Breitbach:2018ddu,Pi:2019ihn,Khlopov:2023mpo,Huang:2015gka,Huang:2015gca,Huang:2017gmp}
and the recent review Ref. \cite{Odintsov:2022cbm}.

Before getting to the core of this work, it is worth recalling the
expression for the energy spectrum of the primordial gravitational
waves for the general relativistic case, assuming a primordial
inflationary era and a dark energy late-time era. We shall
consider a spatially flat Friedmann-Robertson-Walker (FRW)
background with metric,
\begin{equation}
\label{metric} \centering {\rm d}s^2=-{\rm
d}t^2+a(t)^2\sum_{i=1}^{3}{({\rm d} x^{i})^2}\, ,
\end{equation}
with $a(t)$ denoting the scale factor and $t$ stands for the
cosmic time. Since we are interested in inflationary originating
gravitational waves, we shall use the conformal time $\tau$, with
the FRW line element written in this case as follows,
\begin{equation}
  {\rm d}s^{2}=a^{2}[-{\rm d}\tau^{2}+(\delta_{ij}+h_{ij})
  {\rm d}x^{i}{\rm d}x^{j}],
\end{equation}
and the $x^{i}$ stand for the comoving spatial coordinates, while
the tensor $h_{ij}$ denotes the gauge-invariant tensor
perturbation of the metric, which is symmetric
($h_{ij}\!=\!h_{ji}$), traceless $h_{ii}\!=\!0$, and transverse
$\partial^j h_{ij}\!=\!0$. Now if the tensor perturbation is
considered as a quantum field $h_{ij}(\tau,{\bf x})$ as a quantum
field in the unperturbed FRW spacetime $g_{\mu\nu}
=diag\{-a^{2},a^{2},a^{2},a^{2}\}$, we can construct the second
order quadratic action of the tensor perturbations,
\begin{equation}
  \label{tensor_action}
  S=\int d\tau d{\bf x}\sqrt{-g}\left[
    \frac{-g^{\mu\nu}}{64\pi G}\partial_{\mu}h_{ij}
    \partial_{\nu}h_{ij}+\frac{1}{2}\Pi_{ij}h_{ij}\right].
\end{equation}
where $g^{\mu\nu}$ is the inverse of $g_{\mu\nu}$ and also $g$
denotes the determinant of the metric as usual. The tensor
$\Pi_{\mu \nu}$ stands for the anisotropic stress $\Pi_{\mu \nu}$,
which is defined as,
\begin{equation}
  \Pi^{i}_{j}=T^{i}_{j}-p\delta^{i}_{j}
\end{equation}
and also satisfies $\Pi_{ii}=0$ and also the transverse condition
$\partial^{i}\Pi_{ij}=0$. Note that it is coupled to the metric
tensor perturbation $h_{ij}$ and thus it acts as a source in the
gravitational quantum action (\ref{tensor_action}). By varying the
gravitational quantum action (\ref{tensor_action}) with respect to
the $h_{ij}$, we obtain the evolution equation of the metric
perturbation,
\begin{equation}
  \label{h_eq}
  h_{ij}''+2\frac{a'(\tau)}{a(\tau)}h_{ij}'-{\bf \nabla}^{2}h_{ij}
  =16\pi G a^{2}(\tau)\Pi_{ij}(\tau,{\bf x}),
\end{equation}
with the prime denoting differentiation with respect to the
conformal time $\tau$. Taking the fourier transformation of the
evolution equation (\ref{h_eq}), we get,
\begin{subequations}
  \label{fourier_expand}
  \begin{eqnarray}
    h_{ij}^{}(\tau,{\bf x})\!\!&=\!&\!\!\sum_{r}
    \sqrt{16\pi G}\!\!\int\!\!\!
    \frac{d{\bf k}}{(2\pi)^{3/2}}\epsilon_{ij}^{r}({\bf k})
    h_{{\bf k}}^{r}(\tau){\rm e}^{i{\bf k}{\bf x}},\qquad\quad \\
    \Pi_{ij}^{}(\tau,{\bf x})\!\!&=\!&\!\!\sum_{r}
    \sqrt{16\pi G}\!\!\int\!\!\!
    \frac{d{\bf k}}{(2\pi)^{3/2}}\epsilon_{ij}^{r}({\bf k})
    \Pi_{{\bf k}}^{r}(\tau){\rm e}^{i{\bf k}{\bf x}},\qquad\quad
  \end{eqnarray}
\end{subequations}
with $r=$(``$+$'' or ``$\times$'') indicating the polarization of
the gravitational tensor perturbation. The polarization tensors
satisfy [$\epsilon_{ij}^{r} ({\bf
  k})=\epsilon_{ji}^{r}({\bf k})$], and also are traceless $\epsilon_{ii}^{r}
({\bf k})=0$, and transverse $k_{i}\epsilon_{ij}^{r}({\bf k})=0$.
We shall choose a circular polarization basis satisfying
$\epsilon_{ij}^{r}({\bf k})=(\epsilon_{ij}^{r} (\textrm{-}{\bf
  k}))^{\ast}$, so the polarization basis can be normalized as follows,
\begin{equation}
  \label{basis_norm}
  \sum_{i,j}\epsilon_{ij}^{r}({\bf k})(\epsilon_{ij}^{s}
  ({\bf k}))^{\ast}=2\delta^{rs}.
\end{equation}
Eq. (\ref{fourier_expand}) in conjunction with Eq.
(\ref{tensor_action}), yields,
\begin{equation}
  \label{tensor_action_fourier}
  S\!=\!\!\sum_{r}\!\!\int\!\!d\tau d{\bf k}\frac{a^{2}}{2}\;\!\!
  \Big[h_{{\bf k}}^{r}{}'h_{\!\textrm{-}{\bf k}}^{r}\!{}'\!
  -\!k^{2}h_{{\bf k}}^{r}h_{\!\textrm{-}{\bf k}}^{r}\!
  +\!32\pi G a^{2}\Pi_{{\bf k}}^{r}h_{\!\textrm{-}{\bf k}}^{r}
  \Big]\, ,
\end{equation}
which is nothing but the tensor perturbations action
(\ref{tensor_action}), written in terms of the Fourier
transformation of the metric perturbation. Now, imposing the
canonical quantization conditions to the above action, using as a
canonical configuration space variable the perturbation $h_{{\bf
k}}^{r}$ and the conjugate momentum being,
\begin{equation}
  \label{def_pi}
  \pi_{{\bf k}}^{r}(\tau)=a^{2}(\tau)h_{\!\textrm{-}{\bf
  k}}^{r}{}'(\tau)\, ,
\end{equation}
therefore by making these quantum operators $\hat{h}_{{\bf
k}}^{r}$ and $\hat{\pi}_{{\bf k}}^{r}$, the theory at hand can be
quantized by using the following equal time commutation relation,
\begin{subequations}
  \label{h_pi_commutators}
  \begin{eqnarray}
    \left[\hat{h}_{{\bf k}}^{r}(\tau),
      \hat{\pi}_{{\bf k}'}^{s}(\tau)\right]
    &=&i\delta^{rs}\delta^{(3)}({\bf k}-{\bf k}'), \\
    \left[\hat{h}_{{\bf k}}^{r}(\tau),
      \hat{h}_{{\bf k}'}^{s}(\tau)\right]
    &=&\left[\hat{\pi}_{{\bf k}}^{r}(\tau),
      \hat{\pi}_{{\bf k}'}^{s}(\tau)\right]=0\, ,
  \end{eqnarray}
\end{subequations}
and the Fourier components of $\hat{h}_{ij}(\tau,{\bf x})$ satisfy
the following relation $\hat{h}_{{\bf
k}}^{r}=\hat{h}_{\!\textrm{-}{\bf k}}^{r\dag}$, since
$\hat{h}_{ij}(\tau,{\bf x})$ is a Hermitian operator. Accordingly
we have,
\begin{equation}
  \label{h_from_a}
  \hat{h}_{{\bf k}}^{r}(\tau)=h_{k}^{}(\tau)\hat{a}_{{\bf k}}^{r}
  +h_{k}^{\ast}(\tau)\hat{a}_{\!\textrm{-}{\bf k}}^{r\dag},
\end{equation}
and the $\hat{a}_{{\bf k}}^{r\dag}$ are the creation operators,
while $\hat{a}_{{\bf k}}^{r}$ are the annihilation operators,
which obey,
\begin{subequations}
  \label{a_commutators}
  \begin{eqnarray}
    \Big[\hat{a}_{{\bf k}}^{r},\hat{a}_{{\bf k}'}^{s\dag}\Big]
    &=&\delta^{rs}\delta^{(3)}({\bf k}-{\bf k}'), \\
    \Big[\hat{a}_{{\bf k}}^{r},\hat{a}_{{\bf k}'}^{s}\Big]
    &=&\Big[\hat{a}_{{\bf k}}^{r\dag},\hat{a}_{{\bf
    k}'}^{s\dag}\Big]=0\, .
  \end{eqnarray}
\end{subequations}
Note that the linearly-independent modes $h_{k}(\tau)$ and
$h_{k}^{\ast}(\tau)$ satisfy the Fourier transformed evolution
equation of the tensor perturbation,
\begin{equation}
  \label{h_eq_ft}
  h_{k}''+2\frac{a'(\tau)}{a(\tau)}h_{k}'+k^{2}h_{k}^{}=
    16\pi G a^{2}(\tau)\Pi_{k}^{}(\tau).
\end{equation}
By looking Eq.\ (\ref{h_from_a}) it is obvious that the modes
$h_{k}^{}(\tau)$ have an explicit dependence on the conformal time
and on the wavenumber $k=|{\bf k}|$. The Wronskian chronological
past normalization condition,
\begin{equation}
  \label{Wronskian}
  h_{k}^{}(\tau)h_{k}^{\ast}{}'(\tau)-h_{k}^{\ast}(\tau)h_{k}'(\tau)
  =\frac{i}{a^{2}(\tau)}
\end{equation}
basically is the compatibility constraint between Eqs.
(\ref{h_pi_commutators}) and (\ref{a_commutators}). An essential
feature of the standard approach for the primordial gravitational
waves is the initial condition chosen for the Fourier mode of the
tensor perturbation in the chronological past which assumes a
Bunch-Davies initial vacuum state,
\begin{equation}
  \label{h_bc}
  h_{k}^{}(\tau)\to\frac{{\rm exp}(-ik\tau)}{a(\tau)\sqrt{2k}}
  \qquad({\rm as}\;\;\tau\to-\infty),
\end{equation}
describing subhorizon modes $\vec{k}$ during inflation. Now let us
proceed to the calculation of the energy spectrum of the
gravitational waves $\Omega_{gw}^{}(k,\tau)$, so an essential
quantity is the $\Delta_{h}^{2}(k,\tau)$, which can be defined by
the following,
\begin{equation}
  \langle0|\hat{h}_{ij}^{}(\tau,{\bf x})\hat{h}_{ij}^{}
  (\tau,{\bf x})|0\rangle\!=\!\!\!\int_{0}^{\infty}\!\!\!\!\!64\pi G
  \frac{k^{3}}{2\pi^{2}}\!\left|h_{k}^{}(\tau)\right|^{2}
  \!\frac{dk}{k},
\end{equation}
and is explicitly,
\begin{equation}
  \label{def_tensor_power}
  \Delta_{h}^{2}(k,\tau)\equiv\frac{d\langle0|\hat{h}_{ij}^{2}
    |0\rangle}{d\,{\rm ln}\,k}=64\pi G\frac{k^{3}}{2\pi^{2}}
  \left|h_{k}^{}(\tau)\right|^{2}.
\end{equation}
Then, the energy spectrum of the stochastic primordial
gravitational waves $\Omega_{gw}^{}(k,\tau)$ is defined as,
\begin{equation}
  \label{def_Omega_gw}
  \Omega_{gw}^{}(k,\tau)\equiv\frac{1}{\rho_{crit}^{}(\tau)}
  \frac{d\langle 0|\hat{\rho}_{gw}^{}(\tau)|0\rangle}{d\,{\rm
  ln}\,k}\, ,
\end{equation}
and it essentially is the gravitational-wave energy density
($\rho_{gw}^{}$) per logarithmic wavenumber interval, with the
critical energy density being defined as follows,
\begin{equation}
  \label{def_rho_crit}
  \rho_{crit}^{}(\tau)=\frac{3H^{2}(\tau)}{8\pi G}.
\end{equation}
The stress energy tensor of the tensor perturbation $h_{ij}$ is
calculated by the action (\ref{tensor_action}) and it reads,
\begin{equation}
  \label{T_alpha_beta}
  T_{\alpha\beta}=-2\frac{\delta L}{\delta\bar{g}^{\alpha\beta}}
  +\bar{g}_{\alpha\beta}L,
\end{equation}
with $L$ standing  for the Lagrangian in (\ref{tensor_action}).
Without taking into account dissipation effects quantified by the
anisotropic stress couplings, the gravitational wave energy
density reads,
\begin{equation}
  \label{rho_gw}
  \rho_{gw}^{}=-T_{0}^{0}=\frac{1}{64\pi G}
  \frac{(h_{ij}')^{2}+(\vec{{\bf \nabla}}h_{ij})^{2}}{a^{2}},
\end{equation}
and the corresponding vacuum expectation value is,
\begin{equation}
  \label{rho_gw_expect}
  \langle0|\rho_{gw}^{}|0\rangle=\int_{0}^{\infty}\frac{k^{3}}
  {2\pi^{2}}\frac{\left|h_{k}'\right|^{2}
    +k^{2}\left|h_{k}^{}\right|^{2}}{a^{2}}\frac{dk}{k},
\end{equation}
and finally the stochastic gravitational wave energy spectrum is,
\begin{equation}
  \Omega_{gw}^{}(k,\tau)=\frac{8\pi G}{3H^{2}(\tau)}
  \frac{k^{3}}{2\pi^{2}}
  \frac{\left|h_{k}'(\tau)\right|^{2}
    +k^{2}\left|h_{k}^{}(\tau)\right|^{2}}{a^{2}(\tau)}.
\end{equation}
We can use $|h_{k}'(\tau)|^{2}=k^{2}|h_{k}^{}(\tau)|^{2}$, thus
the present day gravitational wave energy spectrum is,
\begin{equation}
  \label{spec_relations}
  \Omega_{gw}(k,\tau)=\frac{1}{12}\frac{k^{2}\Delta_{h}^{2}(k,\tau)}
  {H_0^{2}(\tau)}\, ,
\end{equation}
and note that we made the assumption that the present day scale
factor is unity, in order for comoving scales to be identical with
the physical ones, while $H_0$ is the present day Hubble rate. An
important issue to note is that the amplitude of the gravity wave
with comoving wavenumber $k$ is basically frozen when the mode is
superhorizon, however once it renters the Hubble horizon, the
amplitude is damped. Considering modes reentering the Hubble
horizon during matter domination we get,
\begin{equation}
    h_k^{\lambda}(\tau)=h_k^{\lambda ({\rm p})}
    \left( \frac{3j_1(k\tau)}{k\tau}\right),
\end{equation}
with $j_{\ell}$ standing for the $\ell$-th spherical Bessel
function. Assuming a power-law evolution $a(t)\propto t^p$, the
Fourier transformed tensor perturbation has the following form,
\begin{equation}
    h_k(\tau) \propto
    a(t)^{\frac{1-3p}{2p}}J_{\frac{3p-1}{2(1-p)}}( k\tau ),
\end{equation}
with $J_n(x)$ being the Bessel function. It is also vital to take
into account another damping effect in the early Universe, caused
by the fact that the relativistic degrees of freedom are not
constant and the overall scale factor actually behaves as $a(t)
\propto T^{-1}$ \cite{Watanabe:2006qe}, thus the total damping
factor has the form,
\begin{equation}
    \left ( \frac{g_*(T_{\rm in})}{g_{*0}} \right )
    \left ( \frac{g_{*s0}}{g_{*s}(T_{\rm in})} \right )^{4/3},
\end{equation}
where $T_{\rm in}$ stands for the temperature at the horizon
reentry,
\begin{equation}
    T_{\rm in}\simeq 5.8\times 10^6~{\rm GeV}
    \left ( \frac{g_{*s}(T_{\rm in})}{106.75} \right )^{-1/6}
    \left ( \frac{k}{10^{14}~{\rm Mpc^{-1}}} \right ).
\end{equation}
Also, since we assumed that the Universe currently is at the dark
energy era, another damping factor must be taken into account
$\sim (\Omega_m/\Omega_\Lambda)^2$. Thus the present day energy
spectrum of the primordial gravitational waves is,
\begin{equation}
    \Omega_{\rm gw}(f)= \frac{k^2}{12H_0^2}\Delta_h^2(k),
    \label{GWspec}
\end{equation}
where $\Delta_h^2(k)$ is equal to,
\begin{equation}
\Delta_h^2(k)=\Delta_h^{({\rm p})}(k)^{2} \left (
\frac{\Omega_m}{\Omega_\Lambda} \right )^2 \left (
\frac{g_*(T_{\rm in})}{g_{*0}} \right ) \left (
\frac{g_{*s0}}{g_{*s}(T_{\rm in})} \right )^{4/3} \left
(\overline{ \frac{3j_1(k\tau_0)}{k\tau_0} } \right )^2 T_1^2\left
( x_{\rm eq} \right ) T_2^2\left ( x_R \right )\, .
\label{mainfunctionforgravityenergyspectrum}
\end{equation}
Note that the oscillating term must be evaluated over a Hubble
time. Also $\Delta_h^{({\rm p})}(k)^{2}$ denotes the primordial
tensor power spectrum of the inflationary era, which is equal to,
\begin{equation}
\Delta_h^{({\rm
p})}(k)^{2}=\mathcal{A}_T(k_{ref})\left(\frac{k}{k_{ref}}
\right)^{n_{\mathcal{T}}}\, ,
\label{primordialtensorpowerspectrum}
\end{equation}
which must be evaluated at the CMB pivot scale
$k_{ref}=0.002$$\,$Mpc$^{-1}$, while $n_{\mathcal{T}}$ is the
tensor spectral index, and $\mathcal{A}_T(k_{ref})$ is the tensor
perturbations amplitude which  in terms of the amplitude of the
scalar perturbations $\mathcal{P}_{\zeta}(k_{ref})$ is written as
follows,
\begin{equation}\label{amplitudeoftensorperturbations}
\mathcal{A}_T(k_{ref})=r\mathcal{P}_{\zeta}(k_{ref})\, ,
\end{equation}
where $r$ is the tensor-to-scalar ratio. In effect we have,
\begin{equation}\label{primordialtensorspectrum}
\Delta_h^{({\rm
p})}(k)^{2}=r\mathcal{P}_{\zeta}(k_{ref})\left(\frac{k}{k_{ref}}
\right)^{n_{\mathcal{T}}}\, .
\end{equation}
The transfer function $T_1(x_{\rm eq})$ in Eq.
(\ref{mainfunctionforgravityenergyspectrum}) connects the present
day gravitational wave spectrum of the mode $k$ reentering the
Hubble horizon at matter-radiation equality and it is equal to,
\begin{equation}
    T_1^2(x_{\rm eq})=
    \left [1+1.57x_{\rm eq} + 3.42x_{\rm eq}^2 \right ], \label{T1}
\end{equation}
where $x_{\rm eq}=k/k_{\rm eq}$ and $k_{\rm eq}\equiv a(t_{\rm
eq})H(t_{\rm eq}) = 7.1\times 10^{-2} \Omega_m h^2$ Mpc$^{-1}$.
The transfer function $T_2(x_R)$ connects the present day
gravitational wave spectrum of the mode $k$ reentering the Hubble
horizon at reheating and before this was completed, hence for
$k>k_R$, and it is equal to,
\begin{equation}\label{transfer2}
 T_2^2\left ( x_R \right )=\left(1-0.22x^{1.5}+0.65x^2
 \right)^{-1}\, ,
\end{equation}
where $x_R=\frac{k}{k_R}$, and the reheating temperature
wavenumber $k_R$ is,
\begin{equation}
    k_R\simeq 1.7\times 10^{13}~{\rm Mpc^{-1}}
    \left ( \frac{g_{*s}(T_R)}{106.75} \right )^{1/6}
    \left ( \frac{T_R}{10^6~{\rm GeV}} \right )\, ,  \label{k_R}
\end{equation}
with $T_R$ denoting the reheating temperature. The reheating
temperature is a freely chosen variable and will play a crucial
role in the analysis that follows. Also, the expression for
$g_*(T_{\mathrm{in}}(k))$ in Eq.
(\ref{mainfunctionforgravityenergyspectrum}) is
\cite{Kuroyanagi:2014nba},
\begin{align}\label{gstartin}
& g_*(T_{\mathrm{in}}(k))=g_{*0}\left(\frac{A+\tanh \left[-2.5
\log_{10}\left(\frac{k/2\pi}{2.5\times 10^{-12}\mathrm{Hz}}
\right) \right]}{A+1} \right) \left(\frac{B+\tanh \left[-2
\log_{10}\left(\frac{k/2\pi}{6\times 10^{-19}\mathrm{Hz}} \right)
\right]}{B+1} \right)\, ,
\end{align}
with $A$ and $B$ being defined as,
\begin{equation}\label{alphacap}
A=\frac{-1-10.75/g_{*0}}{-1+10.75g_{*0}}\, ,
\end{equation}
\begin{equation}\label{betacap}
B=\frac{-1-g_{max}/10.75}{-1+g_{max}/10.75}\, ,
\end{equation}
where $g_{max}=106.75$ and $g_{*0}=3.36$. Also
$g_{*0}(T_{\mathrm{in}}(k))$ can be determined by using Eqs.
(\ref{gstartin}), (\ref{alphacap}) and (\ref{betacap}), by simply
replacing $g_{*0}=3.36$ with $g_{*s}=3.91$.

Coming back to the focus of this paper, our main assumption is
that the Universe, in the pre-CMB era, is described by a stiff
EoS. Suppose that this occurs at a wavenumber
$k_s=0.9\,$Mpc$^{-1}$ and the total EoS parameter in this era is
$w=1$ or slightly smaller, in the range $w\sim[0.25,1]$. For the
purposes of this paper, we shall mainly assume that $w=1$, but we
shall allow in some cases the total EoS parameter $w$ to take
values in the range $w\sim[0.25-1]$. Thus if the total EoS during
the matter domination era is changed for some reason from $w=0$ to
$w=1$, the $h^2$-scaled energy spectrum of the primordial
gravitational waves acquires a total multiplication factor of the
form $\sim \left(\frac{k}{k_{s}}\right)^{r_s}$, with
$r_s=-2\left(\frac{1-3 w}{1+3 w}\right)$
\cite{Gouttenoire:2021jhk}. Thus the total $h^2$-scaled energy
spectrum of the primordial gravitational waves has the form,
\begin{align}
\label{GWspecfRnewaxiondecay}
    &\Omega_{\rm gw}(f)=S_k(f)\times \frac{k^2}{12H_0^2}r\mathcal{P}_{\zeta}(k_{ref})\left(\frac{k}{k_{ref}}
\right)^{n_{\mathcal{T}}} \left ( \frac{\Omega_m}{\Omega_\Lambda}
\right )^2
    \left ( \frac{g_*(T_{\rm in})}{g_{*0}} \right )
    \left ( \frac{g_{*s0}}{g_{*s}(T_{\rm in})} \right )^{4/3} \nonumber  \left (\overline{ \frac{3j_1(k\tau_0)}{k\tau_0} } \right )^2
    T_1^2\left ( x_{\rm eq} \right )
    T_2^2\left ( x_R \right )\, ,
\end{align}
with $S_k(f)$,
\begin{equation}\label{multiplicationfactor1}
S_k(f)=\left(\frac{k}{k_{s}}\right)^{r_s}\, ,
\end{equation}
with $k_{ref}$ being the CMB pivot scale
$k_{ref}=0.002$$\,$Mpc$^{-1}$ as we mentioned earlier, while
$n_{\mathcal{T}}$ and $r$ denote the tensor spectral index of the
primordial tensor perturbations and the tensor-to-scalar ratio
respectively. Of course as we already mentioned, the reheating
temperature is a variable in our theory, and basically its scale
is still a mystery that refers to the hypothetical reheating era.

What now remains is to model the inflationary era, and we shall
use two types of models, one which yields a red-tilted tensor
spectral index with the standard consistency relation
$n_{\mathcal{T}}=-r/8$ or a general red tilted tensor spectral
index, and one model with a blue-tilted tensor spectral index of
the order $n_{\mathcal{T}}\sim \mathcal{O}(0.37)$. The red-tilted
model can be any of the viable single scalar field theory such as
the $a$-attractors \cite{alpha1,alpha2,alpha3}, or a standard
$R^2$ model. It is worth to quote the essential features of the
aforementioned two theories which yield a red-tilted tensor
spectral index. Consider first the $a$-attractors, which is a very
popular and elegant class of viable models.

These models originate from a non-minimally coupled scalar field
Jordan frame action,
\begin{equation}\label{c1}
\mathcal{S}_J=\int
d^4x\Big{[}f(\phi)R-\frac{\omega(\phi)}{2}g^{\mu
\nu}\partial_{\mu}\phi\partial_{\nu}\phi-U(\phi)\Big{]}\, .
\end{equation}
Upon performing a conformal transformation of the form,
\begin{equation}\label{c4}
\tilde{g}_{\mu \nu}=\Omega^2g_{\mu \nu}\, ,
\end{equation}
the Einstein frame action  can be obtained, with the ``tilde''
hereafter denoting the Einstein frame physical quantities. In
order to obtain the Einstein frame minimal scalar field action we
can make the choice,,
\begin{equation}\label{c6}
\Omega^2=\frac{2}{M_p^2}f(\phi)\, ,
\end{equation}
therefore, after the conformal transformation, the Einstein frame
action is obtained,
\begin{equation}\label{c12}
\mathcal{S}_E=\int
d^4x\sqrt{-\tilde{g}}\Big{[}\frac{M_p^2}{2}\tilde{R}-\frac{\zeta
(\phi)}{2} \tilde{g}^{\mu \nu }\tilde{\partial}_{\mu}\phi
\tilde{\partial}_{\nu}\phi-V(\phi)\Big{]}+S_m(\Omega^{-2}\tilde{g}_{\mu
\nu},\psi_m)\, ,
\end{equation}
and the Einstein frame potential $V(\phi)$ is related to the
Jordan frame scalar potential $U(\phi)$ as follows,
\begin{equation}\label{c13}
V(\phi)=\frac{U(\phi)}{\Omega^4}\, ,
\end{equation}
with $\zeta(\phi)$ being equal to,
\begin{equation}\label{c14}
\zeta
(\phi)=\frac{M_p^2}{2}\Big{(}\frac{3\Big{(}\frac{df}{d\phi}\Big{)}^2}{f^2}+\frac{2\omega(\phi)}{f}\Big{)}\,
.
\end{equation}
The kinetic term can be made canonical by using the following,
\begin{equation}\label{c15}
\Big{(}\frac{d\varphi}{d \phi}\Big{)} =\sqrt{\zeta(\phi)}\, ,
\end{equation}
thus the Einstein frame action reads,
\begin{equation}\label{c17}
\mathcal{S}_E=\int
d^4x\sqrt{-\tilde{g}}\Big{[}\frac{M_p^2}{2}\tilde{R}-\frac{1}{2}\tilde{g}^{\mu
\nu } \tilde{\partial}_{\mu}\varphi
\tilde{\partial}_{\nu}\varphi-V(\varphi)\Big{]}
\end{equation}
with,
\begin{equation}\label{c18}
V(\varphi)=\frac{U(\varphi)}{\Omega^4}=\frac{U(\varphi)}{4
M_p^4f^2}\, .
\end{equation}
The $a$-attractor class of inflationary potentials is obtained
with the choice,
\begin{equation}\label{alphaattrcondition}
\omega (\phi)=\frac{\Big{(}\frac{df}{d\phi}\Big{)}^2}{4\xi f}\, ,
\end{equation}
with $\xi$ being an arbitrary parameter of this class of models.
In terms of (\ref{alphaattrcondition}), Eq. (\ref{c15}) takes the
form,
\begin{equation}\label{c15alpha}
\Big{(}\frac{d\varphi}{d \phi}\Big{)}
=M_p\sqrt{\frac{3a}{2}}\frac{\Big{(}\frac{df}{d\phi}\Big{)}}{ f}\,
,
\end{equation}
with $a$ being defined in the following way,
\begin{equation}\label{alphaprm}
a=1+\frac{1}{6\xi}\, .
\end{equation}
In addition we have,
\begin{equation}\label{c15newalpha}
\varphi=M_p\sqrt{\frac{3\alpha}{2}}\ln f\, ,
\end{equation}
and it is noticeable that the above is valid regardless the choice
of the arbitrary function $f$. One can observe the universality
class of the $a$-attractor potentials already emerging at this
level. By using Eq. (\ref{c15newalpha}), we obtain,
\begin{equation}\label{fasfunctionofphi}
f=e^{\sqrt{\frac{2}{3\alpha }}\frac{\varphi}{M_p}}\, .
\end{equation}
The $a$-attractor class of potentials is obtained if the scalar
potential in the Jordan frame is chosen to be,
\begin{equation}\label{alphaattractorsjordanpot}
U(\phi)=V_0f^2\Big{(}1-\frac{1}{f}\Big{)}^{2n}\, ,
\end{equation}
hence the Einstein frame potential is written as,
\begin{equation}\label{einsteinframepot}
V(\varphi)=\mathcal{V}_0\Big{(}1-\frac{1}{f}\Big{)}^{2n}=\tilde{V}_0M_p^4\Big{(}1-\frac{1}{f}\Big{)}^{2n}\,
,
\end{equation}
The inflationary indices for $a\leq 1$ have the following form,
\begin{equation}\label{spectralindexsmallalpha}
n_s\simeq 1-\frac{2}{N}\, ,\,\,\,r=\frac{12a}{N^2}\, ,
\end{equation}
and $n_{\mathcal{T}}=-r/8$, and this holds true for all the
possible choices of the arbitrary function $f(\phi(\varphi))$.

Equivalently, a red-tilted tensor spectral index can be obtained
by $F(R)$ gravity theories. For completeness we shall demonstrate
in brief the formalism. Consider a vacuum $F(R)$ gravity action,
\begin{equation}\label{action1dse}
\mathcal{S}=\frac{1}{2\kappa^2}\int \mathrm{d}^4x\sqrt{-g}F(R),
\end{equation}
The field equations for a FRW metric are,
\begin{align}
\label{JGRG15} 0 =& -\frac{F(R)}{2} + 3\left(H^2 + \dot H\right)
F_R(R) - 18 \left( 4H^2 \dot H + H \ddot H\right) F_{RR}(R)\, ,\\
\label{Cr4b} 0 =& \frac{F(R)}{2} - \left(\dot H +
3H^2\right)F_R(R) + 6 \left( 8H^2 \dot H + 4 {\dot H}^2 + 6 H
\ddot H + \dddot H\right) F_{RR}(R) + 36\left( 4H\dot H + \ddot
H\right)^2 F_{RRR} \, ,
\end{align}
where $F_{RR}=\frac{\mathrm{d}^2F}{\mathrm{d}R^2}$, and
$F_{RRR}=\frac{\mathrm{d}^3F}{\mathrm{d}R^3}$. We also shall
assume that the slow-roll conditions hold true,
\begin{equation}\label{slowrollconditionshubble}
\ddot{H}\ll H\dot{H},\,\,\, \frac{\dot{H}}{H^2}\ll 1\, .
\end{equation}
The inflationary indices capture the dynamics during inflation,
namely $\epsilon_1$ ,$\epsilon_2$, $\epsilon_3$, $\epsilon_4$,
which are \cite{Hwang:2005hb,reviews1,reviews6},
\begin{equation}
\label{restofparametersfr}\epsilon_1=-\frac{\dot{H}}{H^2}, \quad
\epsilon_2=0\, ,\quad \epsilon_3= \frac{\dot{F}_R}{2HF_R}\, ,\quad
\epsilon_4=\frac{\ddot{F}_R}{H\dot{F}_R}\,
 ,
\end{equation}
and in terms of these, the spectral index of the scalar
perturbations and the tensor-to-scalar ratio are written as
follows \cite{reviews1,Hwang:2005hb},
\begin{equation}
\label{epsilonall} n_s=
1-\frac{4\epsilon_1-2\epsilon_3+2\epsilon_4}{1-\epsilon_1},\quad
r=48\frac{\epsilon_3^2}{(1+\epsilon_3)^2}\, .
\end{equation}
Using the Raychaudhuri equation, we get,
\begin{equation}\label{approx1}
\epsilon_1=-\epsilon_3(1-\epsilon_4)\, ,
\end{equation}
hence in view of the slow-roll conditions we have approximately
$\epsilon_1\simeq -\epsilon_3$, hence, the scalar spectral index
takes the form,
\begin{equation}
\label{spectralfinal} n_s\simeq 1-6\epsilon_1-2\epsilon_4\, ,
\end{equation}
and the tensor-to-scalar ratio takes the form,
\begin{equation}
\label{tensorfinal} r\simeq 48\epsilon_1^2\, .
\end{equation}
Regarding $\epsilon_4=\frac{\ddot{F}_R}{H\dot{F}_R}$ we have,
\begin{equation}\label{epsilon41}
\epsilon_4=\frac{\ddot{F}_R}{H\dot{F}_R}=\frac{\frac{d}{d
t}\left(F_{RR}\dot{R}\right)}{HF_{RR}\dot{R}}=\frac{F_{RRR}\dot{R}^2+F_{RR}\frac{d
(\dot{R})}{d t}}{HF_{RR}\dot{R}}\, ,
\end{equation}
and by using,
\begin{equation}\label{rdot}
\dot{R}=24\dot{H}H+6\ddot{H}\simeq 24H\dot{H}=-24H^3\epsilon_1\, ,
\end{equation}
in conjunction with Eq. (\ref{epsilon41}) we obtain,
\begin{equation}\label{epsilon4final}
\epsilon_4\simeq -\frac{24
F_{RRR}H^2}{F_{RR}}\epsilon_1-3\epsilon_1+\frac{\dot{\epsilon}_1}{H\epsilon_1}\,
.
\end{equation}
Using,
\begin{equation}\label{epsilon1newfiles}
\dot{\epsilon}_1=-\frac{\ddot{H}H^2-2\dot{H}^2H}{H^4}=-\frac{\ddot{H}}{H^2}+\frac{2\dot{H}^2}{H^3}\simeq
2H \epsilon_1^2\, ,
\end{equation}
the slow-roll index $\epsilon_4$ takes the final form,
\begin{equation}\label{finalapproxepsilon4}
\epsilon_4\simeq -\frac{24
F_{RRR}H^2}{F_{RR}}\epsilon_1-\epsilon_1\, .
\end{equation}
Now the tensor spectral index can be easily calculated. We have
\cite{Hwang:2005hb,reviews1,Odintsov:2020thl},
\begin{equation}\label{tensorspectralindexr2gravity}
n_{\mathcal{T}}\simeq -2 (\epsilon_1+\epsilon_3)\, ,
\end{equation}
so by using (\ref{finalapproxepsilon4}) we have,
\begin{equation}\label{tensorspectralindexr2ini}
n_{\mathcal{T}}\simeq -2 \frac{\epsilon_1^2}{1+\epsilon_1}\simeq
-2\epsilon_1^2\, .
\end{equation}
Considering the $R^2$ model, we have,
\begin{equation}\label{r2modeltensorspectralindexfinal}
n_{\mathcal{T}}\simeq -\frac{1}{2N^2}\, ,
\end{equation}
hence for $N=60$ we have $n_{\mathcal{T}}=-0.000138889$,
$n_s\simeq 0.963$ and $r\simeq 0.0033$. We shall use these values
in the analysis that follows.

Now let us turn our focus on the blue-tilted tensor spectrum
theory. Since as we will show, we will need a mild blue-tilted
theory, we shall use a class of viable inflationary models which
yield a mild blue-tilt, while simultaneously are compatible with
the Planck 2018 data and with the GW170817 event. An elegant
example of such a theory is the Einstein-Gauss-Bonnet theory of
gravity, which is essentially an example of a canonical scalar
field theory with string corrections included. The formalism for
rendering such a theory compatible with the GW170817 event was
developed in Refs.
\cite{Oikonomou:2022xoq,Oikonomou:2021kql,Odintsov:2020sqy}. We
shall quote an explicit model that has all the desirable features
required for the analysis that will follow.
\begin{figure}[h!]
\centering
\includegraphics[width=40pc]{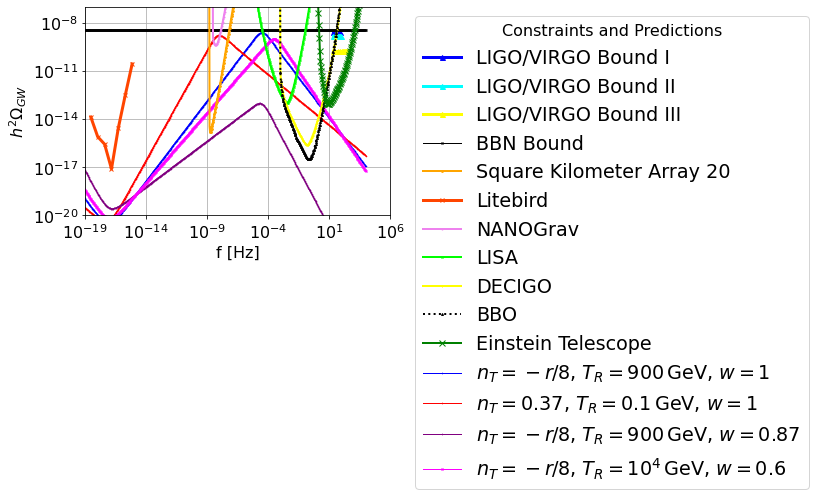}
\caption{The $h^2$-scaled gravitational wave energy spectrum for
the $a$-attractors inflationary model (similar for the $R^2$
model) with $n_{\mathcal{T}}=-r/8$, and for the
Einstein-Gauss-Bonnet model with $n_{\mathcal{T}}=0.37$ for a
stiff era occurring when the modes with $k=0.9\,$Mpc$^{-1}$ enter
the Hubble horizon. The blue curve corresponds to
$n_{\mathcal{T}}=-r/8$, $r=0.003$, $w=1$ and a reheating
temperature $900\,$GeV, the purple curve to
$n_{\mathcal{T}}=-r/8$, $r=0.003$, $w=0.87$ and a reheating
temperature $900\,$GeV, the magenta curve to
$n_{\mathcal{T}}=-r/8$, $r=0.003$, $w=0.6$ and a reheating
temperature $10^4\,$GeV, while the red curve to
$n_{\mathcal{T}}=0.37$, $r=0.003$, $w=1$ and a reheating
temperature $0.1\,$GeV.} \label{plot1}
\end{figure}
The gravitational action of such a theory is,
\begin{equation}
\label{action} \centering
S=\int{d^4x\sqrt{-g}\left(\frac{R}{2\kappa^2}-\frac{1}{2}\partial_{\mu}\phi\partial^{\mu}\phi-V(\phi)-\frac{1}{2}\xi(\phi)\mathcal{G}\right)}\,
,
\end{equation}
with  $\mathcal{G}$ denoting the Gauss-Bonnet invariant which
defined explicitly
$\mathcal{G}=R^2-4R_{\alpha\beta}R^{\alpha\beta}+R_{\alpha\beta\gamma\delta}R^{\alpha\beta\gamma\delta}$
with $R_{\alpha\beta}$ and $R_{\alpha\beta\gamma\delta}$ denoting
as usual the Ricci and Riemann tensors respectively. The speed of
the tensor perturbations is demanded to be equal to unity in such
theories, thus using this constraint, the slow-roll indices for
this theory take the form \cite{Oikonomou:2021kql},
\begin{equation}
\label{index1} \centering \epsilon_1\simeq\frac{\kappa^2
}{2}\left(\frac{\xi'}{\xi''}\right)^2\, ,
\end{equation}
\begin{equation}
\label{index2} \centering
\epsilon_2\simeq1-\epsilon_1-\frac{\xi'\xi'''}{\xi''^2}\, ,
\end{equation}
\begin{equation}
\label{index3} \centering \epsilon_3=0\, ,
\end{equation}
\begin{equation}
\label{index4} \centering
\epsilon_4\simeq\frac{\xi'}{2\xi''}\frac{\mathcal{E}'}{\mathcal{E}}\,
,
\end{equation}
\begin{equation}
\label{index5} \centering
\epsilon_5\simeq-\frac{\epsilon_1}{\lambda}\, ,
\end{equation}
\begin{equation}
\label{index6} \centering \epsilon_6\simeq
\epsilon_5(1-\epsilon_1)\, ,
\end{equation}
where $\mathcal{E}=\mathcal{E}(\phi)$ and also
$\lambda=\lambda(\phi)$ are,
\begin{equation}\label{functionE}
\mathcal{E}(\phi)=\frac{1}{\kappa^2}\left(
1+72\frac{\epsilon_1^2}{\lambda^2} \right),\,\, \,
\lambda(\phi)=\frac{3}{4\xi''\kappa^2 V}\, .
\end{equation}
Accordingly, the observational indices of inflation can be
obtained, which are,
\begin{equation}
\label{spectralindex} \centering
n_{\mathcal{S}}=1-4\epsilon_1-2\epsilon_2-2\epsilon_4\, ,
\end{equation}
\begin{equation}\label{tensorspectralindex}
n_{\mathcal{T}}=-2\left( \epsilon_1+\epsilon_6 \right)\, ,
\end{equation}
\begin{equation}\label{tensortoscalar}
r=16\left|\left(\frac{\kappa^2Q_e}{4H}-\epsilon_1\right)\frac{2c_A^3}{2+\kappa^2Q_b}\right|\,
,
\end{equation}
and $c_A$ denotes the sound speed of the scalar perturbations,
which is not trivial and has the following form,
\begin{equation}
\label{sound} \centering c_A^2=1+\frac{Q_aQ_e}{3Q_a^2+
\dot\phi^2(\frac{2}{\kappa^2}+Q_b)}\, ,
\end{equation}
with,
\begin{align}\label{qis}
& Q_a=-4 \dot\xi H^2,\,\,\,Q_b=-8 \dot\xi H,\,\,\,
Q_t=F+\frac{Q_b}{2},\\
\notag &  Q_c=0,\,\,\,Q_e=-16 \dot{\xi}\dot{H}\, .
\end{align}
In effect, the tensor-to-scalar ratio and the tensor spectral
index take the following form,
\begin{equation}\label{tensortoscalarratiofinal}
r\simeq 16\epsilon_1\, ,
\end{equation}
\begin{equation}\label{tensorspectralindexfinal}
n_{\mathcal{T}}\simeq -2\epsilon_1\left ( 1-\frac{1}{\lambda
}+\frac{\epsilon_1}{\lambda}\right)\, .
\end{equation}
We shall use a successful viable model that fits our constraints
and the purposes of this analysis, which has the following
Gauss-Bonnet coupling function,
\begin{equation}
\label{modelA} \xi(\phi)=\beta  \exp \left(\left(\frac{\phi
}{M}\right)^2\right)\, ,
\end{equation}
with $\beta$ being a dimensionless parameter and $M$ has mass
dimensions $[m]^1$. The novel feature of the formalism of Ref.
\cite{Oikonomou:2021kql} is that in order for the theory to be
compatible with the GW170817 event, the scalar field potential and
the Gauss-Bonnet scalar coupling are not arbitrarily chosen, but
are constrained to satisfy a non-trivial differential equation,
thus by using the form of the Gauss-Bonnet scalar coupling given
in Eq. (\ref{modelA}), the scalar potential is found to have this
form,
\begin{equation}
\label{potA} \centering V(\phi)=\frac{3}{3 \gamma  \kappa ^4+4
\beta  \kappa ^4 e^{\frac{\phi ^2}{M^2}}} \, ,
\end{equation}
with $\gamma$ being an integration constant which is
dimensionless. Thus using the above formalism, the slow-roll
indices in terms of the scalar field easily follow,
\begin{equation}
\label{index1A} \centering \epsilon_1\simeq \frac{\kappa ^2 M^4
\phi ^2}{2 \left(M^2+2 \phi ^2\right)^2} \, ,
\end{equation}
\begin{equation}
\label{index2A} \centering \epsilon_2\simeq \frac{M^4
\left(2-\kappa ^2 \phi ^2\right)-4 M^2 \phi ^2}{2 \left(M^2+2 \phi
^2\right)^2}\, ,
\end{equation}
\begin{equation}
\label{index3A} \centering \epsilon_3=0\, ,
\end{equation}
\begin{equation}
\label{index5A} \centering \epsilon_5\simeq -\frac{4 \beta  \phi
^2 e^{\frac{\phi ^2}{M^2}}}{\left(M^2+2 \phi ^2\right) \left(3
\gamma +4 \beta  e^{\frac{\phi ^2}{M^2}}\right)} \, ,
\end{equation}
\begin{equation}
\label{index6A} \centering \epsilon_6\simeq -\frac{2 \beta  \phi
^2 e^{\frac{\phi ^2}{M^2}} \left(M^4 \left(2-\kappa ^2 \phi
^2\right)+8 M^2 \phi ^2+8 \phi ^4\right)}{\left(M^2+2 \phi
^2\right)^3 \left(3 \gamma +4 \beta  e^{\frac{\phi
^2}{M^2}}\right)} \, ,
\end{equation}
and accordingly, the scalar spectral index, the tensor spectral
index and the tensor-to-scalar ratio are,
\begin{align}\label{spectralpowerlawmodel}
& n_{\mathcal{S}}\simeq -1-\frac{\kappa ^2 M^4 \phi
^2}{\left(M^2+2 \phi ^2\right)^2}+\frac{4 \phi ^2 \left(3 M^2+2
\phi ^2\right)}{\left(M^2+2 \phi ^2\right)^2}\\ & \notag
+\frac{4608 \beta ^2 \phi ^6 e^{\frac{2 \phi ^2}{M^2}} \left(6
\gamma  \phi ^2+16 \beta  e^{\frac{\phi ^2}{M^2}} \left(M^2+\phi
^2\right)+9 \gamma  M^2\right)}{\left(M^2+2 \phi ^2\right)^4
\left(3 \gamma +4 \beta  e^{\frac{\phi ^2}{M^2}}\right)^3} \, ,
\end{align}
regarding the spectral index of scalar perturbations, while
spectral index of tensor perturbations reads,
\begin{align}\label{tensorspectralindexpowerlawmodel}
& n_{\mathcal{T}}\simeq \frac{\phi ^2 \left(-4 \beta e^{\frac{\phi
^2}{M^2}} \left(M^4 \left(3 \kappa ^2 \phi ^2-2\right)+\kappa ^2
M^6-8 M^2 \phi ^2-8 \phi ^4\right)-3 \gamma \kappa ^2 M^4
\left(M^2+2 \phi ^2\right)\right)}{\left(M^2+2 \phi ^2\right)^3
\left(3 \gamma +4 \beta  e^{\frac{\phi ^2}{M^2}}\right)}
 \, ,
\end{align}
\begin{equation}\label{tensortoscalarfinalmodelpowerlaw}
r\simeq \frac{8 \kappa ^2 M^4 \phi ^2}{\left(M^2+2 \phi
^2\right)^2}\, .
\end{equation}
For our analysis we shall need mildly blue tensor spectral index,
so the theory at hand yields such mild values in the range
$n_{\mathcal{T}}=[0.378,0.379]$ with a tensor-to-scalar ratio
$r\sim 0.003$ when we choose the free variables as follows,
$\mu=[22.0914,22.09147]$, $\beta=-1.5$, $\gamma=2$, for
approximately $N=60$ $e$-foldings.

After describing the theoretical framework we shall need for our
analysis of the energy spectrum of the primordial gravitational
waves, let us proceed to the analysis of the energy spectrum of
the primordial gravitational waves and the predictions of our
approach. We shall examine thoroughly both the theoretical
frameworks with blue-tilted and the red-tilted tensor spectrum we
analyzed previously. In Fig. \ref{plot1} we plot the $h^2$-scaled
gravitational wave energy spectrum for the $a$-attractors
inflationary model with $n_{\mathcal{T}}=-r/8$, and for the
Einstein-Gauss-Bonnet model with $n_{\mathcal{T}}=0.37$ for a
stiff era occurring when the modes with $k=0.9\,$Mpc$^{-1}$ enter
the Hubble horizon. The blue curve corresponds to
$n_{\mathcal{T}}=-r/8$, $r=0.003$, $w=1$, $T_R=900\,$GeV and a
reheating temperature $900\,$GeV, the purple curve to
$n_{\mathcal{T}}=-r/8$, $r=0.003$, $w=0.87$, $T_R=900\,$GeV and a
reheating temperature $900\,$GeV, the magenta curve to
$n_{\mathcal{T}}=-r/8$, $r=0.003$, $w=0.6$, $T_R=10^4\,$GeV and a
reheating temperature $10^4\,$GeV, while the red curve to
$n_{\mathcal{T}}=0.37$, $r=0.003$, $w=1$, $T_R=0.1\,$GeV and a
reheating temperature $0.1\,$GeV. Similar results are obtained for
the $R^2$ model which we omit for brevity. As it can be seen in
Fig. \ref{plot1} the NANOGrav 2023 signal can be explained by the
blue-tilted Einstein-Gauss-Bonnet theory we developed previously,
with a relatively mild blue-tilted tensor spectral index
$n_{\mathcal{T}}=0.37$, and with a stiff era occurring before the
CMB and for a reheating temperature of the order $T_R=0.1\,$GeV.
Thus if the stiff era occurred before the CMB, the modes entering
the horizon during that era can significantly affect today's
energy spectrum of the primordial gravitational waves and the
predicted signal can explain the NANOGrav 2023 signal. In,
addition, the same signal can be detected by the SKA, LISA, the
future BBO and DECIGO experiments, but not from the Einstein
Telescope. On the other hand, if the inflationary theory is not a
blue-tilted one and it is described by the $a$-attractors for
example, and a stiff era again occurs before the CMB, the NANOGrav
signal cannot be explained by a stiff era, and it can remain
undetectable by all the experiments for  $w=0.87$, $T_R=900\,$GeV,
but for $w=0.6$, $T_R=10^4\,$GeV and $w=1$, $T_R=900\,$GeV it can
be detected by SKA, LISA, BBO and the DECIGO experiments but not
from the Einstein telescope. It is obvious that the synergy
between gravitational waves experiments is vital towards
understanding the underlying theory that governs the primordial
gravitational waves physics. For example the non-detection by the
Einstein Telescope may favor the scenarios we develop in this
paper and other similar ones. In conclusion, a stiff era before
the CMB can enhance the energy spectrum of the gravitational waves
significantly. How about having several deformations of the
background EoS from reheating up to the CMB scales. Imagine that
several deformations may occur and so the modes entering the
horizon at different eras, may affect the energy spectrum. We
shall discuss such scenario in the following section.

\section{Higgs-Axion Higher Order Couplings and Multiple Post-Electroweak Breaking Axion Stiff Eras}

So far we did not consider a specific model that may yield stiff
deformations of the total EoS of the Universe slightly before the
recombination era, since we used an agnostic approach for this
procedure. Now we shall propose a model that may explain such a
total EoS deformation. This model is based on a Higgs-axion model
with higher order non-renormalizable couplings, which after the
electroweak breaking occurs, affect drastically the axion
potential as we explain shortly. This model is based on the fact
that the electroweak breaking actually occurs at some point in the
past. In all the cases in the literature, the electroweak symmetry
breaking occurs in terms of a thermal first order phase transition
with the symmetry breaking transition being of the order $T\sim
100\,$GeV
\cite{Profumo:2007wc,Damgaard:2013kva,Ashoorioon:2009nf,OConnell:2006rsp,Cline:2012hg,Gonderinger:2012rd,Profumo:2010kp,Gonderinger:2009jp,Barger:2008jx,
Cheung:2012nb,Alanne:2014bra,OConnell:2006rsp,Espinosa:2011ax,Espinosa:2007qk,Barger:2007im,Cline:2013gha,Burgess:2000yq,Kakizaki:2015wua,Cline:2012hg,
Enqvist:2014zqa,Chala:2018ari,Noble:2007kk,Katz:2014bha}. As we
saw in the previous section, in order for the NANOGrav signal to
be explained by an inflationary theory in conjunction with a stiff
deformation of the total EoS parameter before the recombination
era, a low reheating temperature is required, of the order $T\sim
0.1\,$GeV. Now, this feature can put the electroweak symmetry
breaking phase transition in peril of existence if it occurs
thermally. However there exist other ways to break the electroweak
symmetry, that do not invoke a thermal phase transition. We are
developing such a theory currently the details of which will be
given elsewhere \cite{futurework}. For the moment we shall assume
that a high reheating temperature is not required for breaking the
electroweak symmetry and that this breaks non-thermally at some
point during the inflationary era.
\begin{figure}[h!]
\centering
\includegraphics[width=40pc]{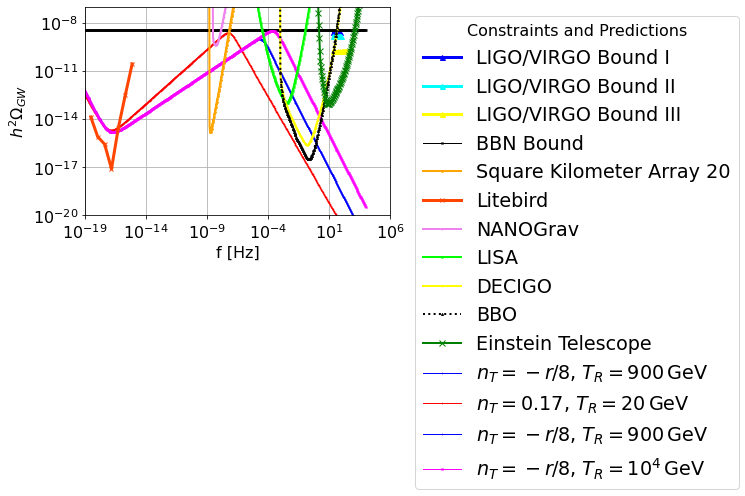}
\caption{The $h^2$-scaled gravitational wave energy spectrum for
the $a$-attractors inflationary model (similar for the $R^2$
model) with $n_{\mathcal{T}}=-r/8$, and for the
Einstein-Gauss-Bonnet model with $n_{\mathcal{T}}=0.17$ for
several deformations of the total background EoS caused by the
axion movements towards the new minimum of its potential at the
post-electroweak symmetry breaking epoch. These deformations in
this scenario occur and affect the modes with wavenumbers
$k_{w_1}=0.001$Mpc$^{-1}$, $k_{w_2}=0.004$Mpc$^{-1}$,
$k_{w_3}=0.9$Mpc$^{-1}$, $k_{t_1}=15\times 10^{5}$Mpc$^{-1}$ ,
$k_{t_2}=10\times 10^{9}$Mpc$^{-1}$, $k_{t_3}=10\times
10^{10}$Mpc$^{-1}$. For those distinct epochs we shall assume that
the total background EoS parameter has the values $w_c=0.34$ for
$k_{w_1}$, $k_{w_2}$ and $k_{t_2}$, $w_{c_1}=1$ for the modes
$k_{w_3}$, $w_b=0.25$ for the modes $k_{t_1}$ and $k_{t_3}$. The
blue curve corresponds to $n_{\mathcal{T}}=-r/8$, $r=0.003$, and a
reheating temperature $900\,$GeV, the magenta curve to
$n_{\mathcal{T}}=-r/8$, $r=0.003$, and a reheating temperature
$10^4\,$GeV, while the red curve to $n_{\mathcal{T}}=0.17$,
$r=0.003$, and a reheating temperature $20\,$GeV. }\label{plot2}
\end{figure}
Now let us discuss in detail the physics of the Higgs-Axion model
that may explain the deformations of the total EoS of the Universe
after the electroweak symmetry breaking and of course before the
CMB era and even after the recombination era, up to present day.
This model was developed in detail in Ref.
\cite{Oikonomou:2023bah} and we shall briefly discuss the
essential features of this model. It is based on
non-renormalizable higher order operators that couple non-trivial
and holomorphically the Higgs and the axion, and for general
Higgs-axion couplings in other contexts, mainly renormalizable,
see for example \cite{Espinosa:2015eda,Im:2019iwd,Dev:2019njv}.
The latter is one of the main dark matter candidates at present
day
\cite{Preskill:1982cy,Abbott:1982af,Dine:1982ah,Marsh:2015xka,Sikivie:2006ni,Co:2019jts,Co:2020dya,Chen:2022nbb,Oikonomou:2022ela,Roy:2021uye,Tsai:2021irw,Visinelli:2018utg,Oikonomou:2022tux,Odintsov:2020iui,Oikonomou:2020qah,Vagnozzi:2022moj,Banerjee:2021oeu,Machado:2019xuc,Machado:2018nqk,Heinze:2023nfb,DiLuzio:2020wdo,Visinelli:2018utg,Mazde:2022sdx,Lambiase:2022ucu,Ramberg:2019dgi}.
Assuming the misalignment axion scenario, which is based on the
pre-inflationary breaking of the axion Peccei-Quinn $U(1)$. During
inflation, and before the electroweak symmetry breaking, the axion
is misaligned from the minimum of its potential, which can be
approximated by,
\begin{equation}\label{axionpotential}
V_a(\phi )\simeq \frac{1}{2}m_a^2\phi^2\, .
\end{equation}
When the Hubble rate becomes of the order of the axion mass, the
axion commences oscillations and thus redshifts as dark matter. We
modify the axion sector by adding the following higher order
non-renormalizable operators of the axion to the Higgs,
\begin{equation}\label{axioneightsixpotential}
V(\phi,h)=V_a(\phi)-m_H^2|H|^2+\lambda_H|H|^4-\lambda\frac{|H|^2\phi^4}{M^2}+g\frac{|H|^2\phi^6}{M^4}\,
,
\end{equation}
with $V_a(\phi)$ being defined in Eq. (\ref{axionpotential}), and
note that the Higgs field at the pre-electroweak symmetry breaking
era  is $H=\frac{h+i h_1}{\sqrt{2}}$, with $m_H=125$ GeV
\cite{ATLAS:2012yve} and $\lambda_H$ is defined via
$\frac{v}{\sqrt{2}}=\left(\frac{-m_H^2}{\lambda_H}\right)^{\frac{1}{2}}$,
and finally $v$ is the electroweak symmetry breaking scale which
is approximately $v\simeq246\,$GeV. We also take the axion mass to
be $m_a\sim 10^{-10}\,$eV and the effective theory scale $M$ to be
$M=20-100\,$TeV. Also the Wilson coefficients are taken to be
$\lambda\sim \mathcal{O}(10^{-20})$ and $g\sim
\mathcal{O}(10^{-5})$. When the electroweak breaking occurs, the
Higgs field becomes $H=v+\frac{h+i h_1}{\sqrt{2}}$, and hence at
leading order, the axion potential at one-loop zero temperature
takes the form,
\begin{equation}\label{axioneffective68}
\mathcal{V}_a(\phi)=V_a(\phi)-\lambda\frac{v^2\phi^4}{M^2}+g\frac{v^2\phi^6}{M^6}+\frac{m_{eff}^4(\phi)}{64\pi^2}\left(
\ln \left(\frac{m_{eff}^2(\phi)}{\mu^2}\right)-\frac{3}{2}\right)
\, ,
\end{equation}
where $m_{eff}^2(\phi)$,
\begin{equation}\label{axioneffectivemass}
m_{eff}^2(\phi)=\frac{\partial^2 V(\phi,h)}{\partial
\phi^2}=m_a^2-\frac{6 \lambda v^2 \phi^2}{M^2}+\frac{15 g v^2
\phi^4}{M^4}\, ,
\end{equation}
and also $\mu$ denotes  the renormalization scale. In this model,
after the electroweak symmetry breaking, the axion potential
develops a minimum which is lower from the original minimum. Hence
the axion begins rolling towards the new minimum, which can be
done in a rapid or a slow-roll way. Hence, the dark matter nature
of the axion is disrupted at some point, and when the axion is
destabilized, the axion rolls towards the new minimum, in a stiff
way as we shall assume in this paper. Once the minimum is reached,
since the Higgs minimum $(h,\phi)=(v,0)$ is energetically
favorable compared to the axion minimum $(h,\phi)=(0,v_s)$, that
is,
\begin{equation}\label{vacuumdecaycondition}
V_h(h)=(v)\gg V(\phi)=(v_s)\, ,
\end{equation}
therefore, the axion vacuum decays to the Higgs vacuum and the
axion starts again in the origin its oscillations, behaving as
dark matter again. The procedure repeats itself perpetually and
thus we have interchanging eras of dark matter and stiff axion
EoS. Now depending on the total background EoS parameter, the
energy spectrum may be affected significantly. At this point we
shall examine some scenarios and their effect on the energy
spectrum of the primordial gravitational waves. Let us consider a
sequence of deformations of the background EoS due to the axion
moving in a kinetic way to the new minimum of its potential in the
post-electroweak era. Let us suppose that these deformations of
the EoS occurred at several distinct epochs starting from
reheating and up to modes that reentered the Hubble horizon just
before the CMB but also after the CMB occurred. So consider that
the background EoS occurred at epochs in which the modes with
wavenumbers $k_{w_1}=0.001$Mpc$^{-1}$, $k_{w_2}=0.004$Mpc$^{-1}$,
$k_{w_3}=0.9$Mpc$^{-1}$, $k_{t_1}=15\times 10^{5}$Mpc$^{-1}$ ,
$k_{t_2}=10\times 10^{9}$Mpc$^{-1}$, $k_{t_3}=10\times
10^{10}$Mpc$^{-1}$. For those distinct epochs we shall assume that
the total background EoS parameter has the values $w_c=0.34$ for
$k_{w_1}$, $k_{w_2}$ and $k_{t_2}$, $w_{c_1}=1$ for the modes
$k_{w_3}$, $w_b=0.25$ for the modes $k_{t_1}$ and $k_{t_3}$. The
inflationary theory is assumed to be one from the $a$-attractors
or the $R^2$, which have red tilt or the Einstein-Gauss-Bonnet
theory we developed in the previous section, which has a
blue-tilted tensor spectral index.
\begin{figure}[h!]
\centering
\includegraphics[width=40pc]{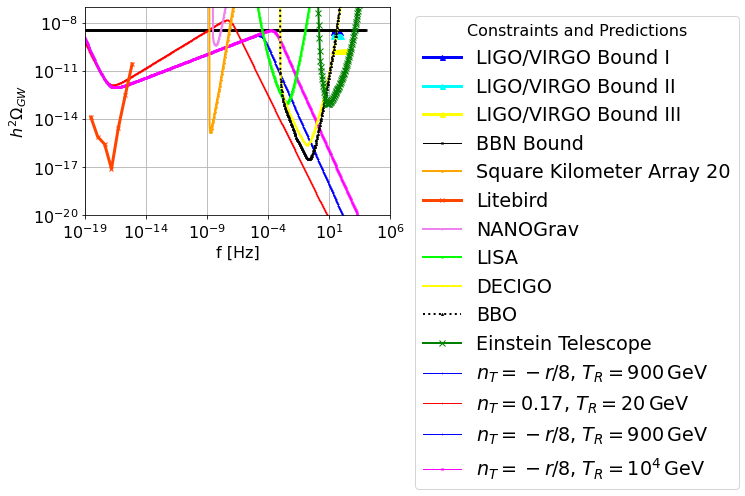}
\caption{The $h^2$-scaled gravitational wave energy spectrum for
the $a$-attractors inflationary model (similar for the $R^2$
model) with $n_{\mathcal{T}}=-r/8$, and for the
Einstein-Gauss-Bonnet model with $n_{\mathcal{T}}=0.17$ for
several deformations of the total background EoS caused by the
axion movements towards the new minimum of its potential at the
post-electroweak symmetry breaking epoch. These deformations in
this scenario occur and affect the modes with wavenumbers
$k_{w_1}=0.001$Mpc$^{-1}$, $k_{w_2}=0.004$Mpc$^{-1}$,
$k_{w_3}=0.9$Mpc$^{-1}$, $k_{t_1}=15\times 10^{5}$Mpc$^{-1}$ ,
$k_{t_2}=10\times 10^{9}$Mpc$^{-1}$, $k_{t_3}=10\times
10^{10}$Mpc$^{-1}$. For those distinct epochs we shall assume that
the total background EoS parameter has the values $w_c=0.34$ for
$k_{w_1}$, $k_{w_2}$ and $k_{t_2}$, $w_{c_1}=1$ for the modes
$k_{w_3}$, $w_b=0.25$ for the mode $k_{t_1}$, and $w=0.2$ for the
mode $k_{t_3}$. The blue curve corresponds to
$n_{\mathcal{T}}=-r/8$, $r=0.003$, and a reheating temperature
$900\,$GeV, the magenta curve to $n_{\mathcal{T}}=-r/8$,
$r=0.003$, and a reheating temperature $10^4\,$GeV, while the red
curve to $n_{\mathcal{T}}=0.17$, $r=0.003$, and a reheating
temperature $20\,$GeV. }\label{plot3}
\end{figure}
For the deformations of the total background EoS parameter, the
energy spectrum is instantaneously modified as follows,
\begin{align}
\label{GWspecfRnewaxiondecay}
    &\Omega_{\rm gw}(f)=S_k(f)\times \frac{k^2}{12H_0^2}r\mathcal{P}_{\zeta}(k_{ref})\left(\frac{k}{k_{ref}}
\right)^{n_{\mathcal{T}}} \left ( \frac{\Omega_m}{\Omega_\Lambda}
\right )^2
    \left ( \frac{g_*(T_{\rm in})}{g_{*0}} \right )
    \left ( \frac{g_{*s0}}{g_{*s}(T_{\rm in})} \right )^{4/3} \nonumber  \left (\overline{ \frac{3j_1(k\tau_0)}{k\tau_0} } \right )^2
    T_1^2\left ( x_{\rm eq} \right )
    T_2^2\left ( x_R \right )\, ,
\end{align}
with the multiplication factor $S_k(f)$ being,
\begin{equation}\label{multiplicationfactor1}
S_k(f)=\left(\frac{k}{k_{t_1}}\right)^{r_{b}}\times
\left(\frac{k}{k_{t_2}}\right)^{r_{c}}\times
\left(\frac{k}{k_{t_3}}\right)^{r_b}\times
\left(\frac{k}{k_{w_1}}\right)^{r_{c}}\times
\left(\frac{k}{k_{w_2}}\right)^{r_{c}}\times
\left(\frac{k}{k_{w_3}}\right)^{r_{c_1}}\, ,
\end{equation}
with $r_b=-2\left(\frac{1-3 w_b}{1+3 w_b}\right)$,
$r_c=-2\left(\frac{1-3 w_c}{1+3 w_c}\right)$ and
$r_{c_1}=-2\left(\frac{1-3 w_{c_1}}{1+3 w_{c_1}}\right)$ and
recall that $w_c=0.34$ for $k_{w_1}$, $k_{w_2}$ and $k_{t_2}$,
$w_{c_1}=1$ for the modes $k_{w_3}$, $w_b=0.25$ for the modes
$k_{t_1}$ and $k_{t_3}$. \cite{Gouttenoire:2021jhk}. In Fig.
\ref{plot2} we present the predicted the energy spectrum of the
primordial gravitational waves for the model at hand for various
reheating temperatures and for the red-tilted tensor and for the
blue-tilted tensor spectral index. The blue curve corresponds to
$n_{\mathcal{T}}=-r/8$, $r=0.003$, and a reheating temperature
$900\,$GeV, the magenta curve to $n_{\mathcal{T}}=-r/8$,
$r=0.003$, and a reheating temperature $10^4\,$GeV, while the red
curve to $n_{\mathcal{T}}=0.17$, $r=0.003$, and a reheating
temperature $20\,$GeV. Thus for the blue-tilted tensor case we
used even milder values for the blue tilt of the tensor spectral
index and higher for the reheating temperature. As it can be seen,
the NANOGrav signal is explained by the Einstein-Gauss-Bonnet
theory with a milder blue tilt in the tensor spectral index,
compared to the values used in the previous section, and a low
reheating temperature, and also the same signal can be detected by
LISA, DECIGO and the BBO detectors, but not from the Einstein
Telescope. Regarding the red-tilted theory, it predicts an energy
spectrum that can be detected by LISA, DECIGO and the BBO detector
but not from the Einstein Telescope. Now the striking new feature
is that in all these cases, the Litebird experiment may detect the
predicted signals. In fact the signal might become stronger if the
era in which the modes with $k_{t_3}=10\times 10^{10}$Mpc$^{-1}$
reentered the Hubble horizon, had a background EoS parameter
$w=0.2$, thus closer to the matter dominated epoch. This can be
seen in Fig. \ref{plot3}. The red curve is incompatible with the
Big Bang Nucleosynthesis constraints, but we just wanted to make a
point here, that the signal can be more pronounced if the modes
with $k_{t_3}=10\times 10^{10}$Mpc$^{-1}$ reentered the Hubble
horizon, had a background EoS parameter $w<0.25$.

Thus the striking new feature of this model is the prediction that
Litebird might detect the tail of this signal, indicating to an
era with background EoS parameter of the order $w=0.2$ for modes
with wavenumber $k\sim \mathcal{O}(10^{10})$Mpc$^{-1}$ which
reentered the horizon during that epoch. Finally, let us mention
again that in order for this model to explain the NANOGrav 2023
signal a low-reheating temperature is required. Now recall that
this model is based on the fact that the electroweak phase
transition takes place primordially, thus with such a
low-reheating temperature, it is impossible to break the
electroweak symmetry thermally. However there exist other ways to
break the electroweak symmetry, apart from the thermal phase
transition. We are developing such a model and will report on this
soon \cite{futurework}.

\subsection{Brief Comment on the Effects of a Short Stiff pre-CMB Era on the BBN and CMB}

Let us here discuss a somewhat important issue regarding the BBN
and CMB effects of a short stiff era that occurred just before the
CMB era and well beyond the matter-radiation equality. As we will
show, since we considered modes with $k=0.9\,$Mpc$^{-1}$, neither
the CMB or the BBN is affected by the short stiff era.

Let us discuss first the BBN issues, which has also been addressed
in \cite{Co:2021lkc}. An even short stiff era could affect the
abundance of light elements, in two ways. If the stiff era
occurred before the BBN, so when the temperature was $T>0.1\,$MeV,
this would affect the helium abundance which is quite sensitive to
the freeze out of proton-neutron conversions. Thus since in our
case the stiff era occurs before the CMB, the abundance of helium
is not affected at all. After the BBN a stiff era would only
affect the abundance of deuterium, which eventually freezes at the
last stages of the BBN. Following \cite{Co:2021lkc}, the
constraint that must be satisfied in order not to  affect the
deuterium abundance is that the temperature at which the kination
era commences must be $T_K<6\,$keV. In our case the stiff era
occurs just before the CMB and well after the matter-radiation
equality, thus the temperature is of the order $\sim 0.5\,$eV or
slightly smaller or even slightly larger. Thus, as it proves there
is no issue related to the abundances of light elements.

Let us discuss the issues that may come up with the CMB. However
even the question of affecting the CMB modes with a stiff era that
occurs when modes of the order $k=0.9\,$Mpc$^{-1}$ (see Fig. 1) is
rather futile. The reason is that modes with $\lambda<10\,$Mpc or
equivalently modes with wavenumber $k>0.628319\,$Mpc$^{-1}$ cannot
be analyzed using linear perturbation theory. Thus since we chose
to study modes with $k=0.9\,$Mpc$^{-1}$, the linear perturbation
theory theory cannot be used and therefore no primary effects are
expected in the CMB polarization, only secondary non-linear
effects. And also it might be possible that these effects might
affect the energy spectrum of primordial gravitational waves by
providing an enhancement of the spectrum. This analysis though
exceeds the aims and scopes of this article, which were to
demonstrate the direct effects of a pre-CMB era on the energy
spectrum of the primordial gravitational waves.

But for completeness, and since at a later section we used modes
with $k<0.628319\,$Mpc$^{-1}$ but we did not take these into
account in the figures, let us qualitatively discuss the effects
of a pre-CMB era on the CMB. The kination era could directly
affect the sound horizon at the last scattering surface and
actually the angular scale of the sound horizon. This is not an
easy problem to address and solve though, since there are quite
many parameters which if their value changes, can compensate the
effects of the stiff era on the sound horizon, for example a
changes in $H_0$, or the baryon fraction and so on. Also the
duration of the stiff era plays a role, which was not considered
\cite{Co:2021lkc} but was analyzed  in \cite{Gouttenoire:2021jhk}.
Hence, the problem is not so easy to solve, and in order to fully
address it the coupled Boltzmann equations which describe the
evolution of matter perturbations and gravitational perturbations
must be solved, and also allow changes in $H_0$, the baryon
fraction and other parameters. This analysis stretches by far
beyond the scopes of this demonstrative article. Besides, our main
assumption was that the stiff era occurred when modes with
wavenumber $k=0.9\,$Mpc$^{-1}$ reentered the Hubble horizon, thus
these modes cannot be treated by using linear perturbation theory.
Hence the whole argument against a stiff era does not apply in our
case.

\section*{Concluding Remarks and Discussion}

In this work we thoroughly examined the quantitative effects of a
stiff era occurring before the CMB era, on the energy spectrum of
the primordial gravitational waves at present day. Firstly, in a
model-agnostic way without giving any reasoning for this kination
era, we examined the effect of this era on the modes that
reentered the horizon during this pre-recombination stiff era. For
our quantitative analysis we focused on modes having wavenumber
$k\sim 0.9\,$Mpc$^{-1}$ and also we assumed some inflationary era,
focusing on viable models. Specifically, the inflationary era can
be realized by a model which generates a red-tilted tensor
spectral index, like the $a$-attractors or the $R^{2}$ model, or
by a model that can generate a blue-tilted tensor spectral index,
like the Einstein-Gauss-Bonnet models. If at the time of the
reentry in the Hubble horizon of the modes with wavenumber $k\sim
0.9\,$Mpc$^{-1}$, the total background EoS parameter describes a
kination era, then the predicted signal can be detected by several
current and future gravitational wave experiments. Specifically,
if the inflationary era has a mild blue-tilted tensor spectral
index of the order $n_{\mathcal{T}}\sim 0.37$ and if the reheating
temperature is of the order $T_R\sim \mathcal{O}(0.1)$GeV, the
signal can explain the 2023 NANOGrav detection of the stochastic
gravitational wave background, and also the signal can be
detectable by the future SKA experiment. Accordingly, if the
reheating temperature is significantly higher and a red-tilted
inflationary is assumed, the predicted energy spectrum will be
detectable by the future LISA, DECIGO and BBO experiments. After
the model-agnostic approach, we considered a specific model
invoking Higgs-axion higher order non-renormalizable couplings.
This model is based on the occurrence of the electroweak symmetry
breaking, which however cannot occur thermally if the reheating
temperature never reached at least 100GeV. There are ways to break
the electroweak symmetry apart from thermal phase transitions
\cite{futurework}, thus for the analysis of this section we
assumed that the electroweak symmetry breaking occurred in a
non-thermal way. Having settled that, we analyzed the Higgs-axion
model which predicts that multiple stiff eras may occur after the
electroweak symmetry breaking, specifically, during reheating,
matter domination era, before and after recombination. These axion
stiff eras changes the background EoS parameter at these eras,
thus affect the modes that reenter the Hubble horizon during these
eras. As we demonstrated, the 2023 NANOGrav stochastic
gravitational wave signal can be explained by this model, if the
reheating temperature is of the order $T_R\sim 20$GeV and the
inflationary era has a really mild blue-tilted tensor spectral
index $n_{\mathcal{T}}\sim 0.17$. The same signal can be
detectable by the SKA. Also for a red-tilted tensor spectral
index, the energy spectrum can be detected by LISA, BBO and
DECIGO. More importantly, in all cases, this model generates a
stochastic signal that can be detected by Litebird, which is a
characteristic feature of this model. In all cases, for both the
model-agnostic approach and for the specific model which we
analyzed, the signal cannot be detectable by the Einstein
Telescope. This brings us to the important line of thinking, that
only the synergy of current and future gravitational wave
experiments can yield decisive information regarding the actual
source of the stochastic gravitational wave background. Patterns
like the ones we described in this paper, will be indicative for
the underlying theory that governs the stochastic gravitational
wave background. For example in the Higgs-axion model, the
blue-tilted tensor spectral index model can be detected from the
future Litebird and the current NANOGrav experiment, and from no
other gravitational wave experiment. This is a characteristic
pattern. Thus the way towards understanding the primordial
Universe will be based on the synergy between current and future
gravitational wave experiments. Exciting upcoming years for
science to say the least.

\section*{Acknowledgments}

This research has been is funded by the Committee of Science of
the Ministry of Education and Science of the Republic of
Kazakhstan (Grant No. AP19674478).

\end{document}